\documentclass{article}
\usepackage[utf8]{inputenc}
\usepackage[round]{natbib}
\usepackage{float}
\usepackage{caption}
\usepackage{nicematrix}
\usepackage{subcaption}
\usepackage{todonotes}
\usepackage{rotating}
\usepackage{amsmath}
\usepackage{algorithm}
\usepackage{algpseudocode}
\usepackage[hidelinks]{hyperref}
\usepackage{pdflscape}
\usepackage{doi}
\usepackage{graphicx} 
\usepackage{booktabs}
\usepackage{amsmath,amsfonts,amssymb,amsthm}
\usepackage{xcolor}
\hypersetup{
    colorlinks,
    linkcolor={red!50!black},
    citecolor={blue!50!black},
    urlcolor={blue!80!black}
}

\usepackage{titling}
\usepackage{lipsum}
\usepackage{multirow}
\usepackage{mathtools}
\usepackage[thinlines]{easytable}
\usepackage{array}
\usepackage[page, toc]{appendix} 
\usepackage{afterpage} 

\usepackage{tikz}
\usepackage{bm}
\usepackage{color, colortbl}
\usepackage{hhline}
\usetikzlibrary{arrows,matrix,positioning,fit, decorations.pathreplacing}
\usepackage{titlesec}

\newcommand{\vecop}[1]{\text{vec}(#1)}

\newcommand{\calL}{\mathcal{L}}

\newcommand{\bA}{\bm A}

\newcommand{\bB}{\bm B}

\newcommand{\bD}{\bm D}
\newcommand{\bd}{\bm d}
\newcommand{\bE}{\bm E}
\newcommand{\be}{\bm e}

\newcommand{\bU}{\bm U}

\newcommand{\bY}{\bm Y}
\newcommand{\by}{\bm y}

\newcommand{\balpha}{\bm \alpha}
\newcommand{\bbeta}{\bm \beta}

\newcommand{\bDelta}{\bm \varDelta}
\newcommand{\bTheta}{\bm \varTheta}

\newcommand{\bPi}{\bm \varPi}
\newcommand{\bSigma}{\bm \varSigma}

\newcommand{\bPhi}{\bm \varPhi}

\theoremstyle{definition}

\definecolor{UMdblue}{RGB}{0,28,61}
\definecolor{UMlblue}{RGB}{0,162,219}
\definecolor{UMorangered}{RGB}{232,78,16}
\definecolor{UMorange}{RGB}{243,148,37} 
\definecolor{UMred}{RGB}{174,11,18} 
\definecolor{UMorangesyllabus}{RGB}{215,92,45}

\definecolor{red1}{RGB}{255,0,0}
\definecolor{red2}{RGB}{250,40,15}
\definecolor{red3}{RGB}{245,0,35}
\definecolor{red4}{RGB}{235,30,30}

\definecolor{blue1}{RGB}{0,0,255}
\definecolor{blue2}{RGB}{40,15,250}
\definecolor{blue3}{RGB}{0,35,245}
\definecolor{blue4}{RGB}{30,30,235}

\definecolor{green1}{RGB}{0,150,0}
\definecolor{green2}{RGB}{15,145,40}
\definecolor{green3}{RGB}{35,145,0}
\definecolor{green4}{RGB}{30,135,30}

\definecolor{yellow1}{RGB}{180,180,20}
\definecolor{yellow2}{RGB}{160,140,30}

\definecolor{ada}{RGB}{247,252,13}
\definecolor{nonada}{RGB}{54,38,134}
\titleformat*{\section}{\large\bfseries}
\titleformat*{\subsection}{\normalsize\bfseries}
\titleformat*{\subsubsection}{\normalsize\bfseries}
\titleformat*{\paragraph}{\normalsize\bfseries}
\titleformat*{\subparagraph}{\normalsize\bfseries}
\usepackage{setspace}
\usepackage{longtable}

\usepackage[left=1in,top=1in,right=1in,bottom=1in,nohead,paperwidth=8.5in, paperheight=11in]{geometry}
\usepackage{authblk}
\theoremstyle{definition}

\theoremstyle{definition}

\theoremstyle{definition}

\theoremstyle{definition}

\theoremstyle{definition}

\usepackage{url}

\usepackage[normalem]{ulem}
\usepackage{color}

\title{\bf Detecting Cointegrating Relations in Non-stationary Matrix-Valued Time Series}
\author{Alain Hecq, Ivan Ricardo\thanks{Corresponding author: Ivan Ricardo, Maastricht University, School of Business and Economics, Department of Quantitative Economics, P.O.Box 616, 6200 MD Maastricht, The Netherlands. E-mail: iu.ricardo@maastrichtuniversity.nl.}, Ines Wilms}
\affil{Maastricht University, Department of Quantitative Economics}
\date{\today}

\begin{document}
\newtheorem{remark}{Remark}

\maketitle

\begin{abstract}
This paper proposes a Matrix Error Correction Model to identify cointegration relations in matrix-valued time series.
We hereby allow separate cointegrating relations along the rows and columns of the matrix-valued time series and use information criteria to select the cointegration ranks.
Through Monte Carlo simulations and a macroeconomic application, we demonstrate that our approach provides a reliable estimation of the number of cointegrating relationships.
\end{abstract}

\bigskip


\newpage
\section{Introduction}

Understanding the long-run relationships between key macroeconomic variables is a central focus for many economists.
Recently, however, interest has turned to cointegration analysis for matrix-valued time series \citep{li2024coint}.
Modeling matrix-valued time series directly allows researchers to capture long-run relationships across multiple dimensions-- such as between countries (row dimension) and economic indicators (column dimension) --providing a more comprehensive understanding of how different economies interact, co-move, and adjust over time.
This paper introduces the Matrix Error Correction Model and demonstrates that information criteria can be reliably used to determine the cointegrating ranks among the different dimensions of the matrix-valued time series.

To fix ideas, we first review the standard framework for analyzing multivariate cointegrated systems.
Let $\by_t$, $t=1, \dots, T$, be an $N$ dimensional time series integrated of order I($1$).
In the presence of cointegration, the vector error correction model (VECM, see e.g., \citealp{johansen1990coint})
\begin{align}
\label{eq:vecm}
\bDelta \by_t = \bd + \balpha \bbeta' \by_{t-1} + \sum_{j=1}^{p} \bPhi_j \bDelta \by_{t-j} + \be_t,
\end{align}
captures long-run relationships, where $\bd$ is the vector of deterministic terms,\footnote{For simplicity, the deterministic terms are left unrestricted and are not included within the cointegrating vector.} and $\bbeta$ and $\balpha$ are the $N \times r$ cointegrating matrix and adjustment coefficients respectively, with $r$ being the cointegrating rank.
Additionally, $\bPhi_j$ represents the $j$th $N \times N$ short-run coefficient matrix, and $\be_t$ is the $N$-dimensional error term.

When $N$ is small-- typically ranging from two to five key variables --the canonical correlation approach of \citet{johansen1991metrica} can reliably determine the cointegration rank of the matrix $\bPi = \balpha \bbeta'$.
However, as $N$ grows, this method becomes less reliable, necessitating alternatives to determine the cointegrating rank (e.g., \citealp{gutierrez2003power}, \citealp{wilms2016forecasting}).
To this end, we exploit the time series' matrix-valued structure, where $N_1$ and $N_2$ denote the number of time series in the rows and columns of the observed matrix over time.
For instance, our empirical analysis examines data from $N_1 = 3$ economic indicators and $N_2 = 4$ countries over $T=116$ observations, resulting in $N_1 N_2 = N = 12$ variables.

This matrix structure allows us to employ a Matrix Error Correction Model (MECM), which jointly accommodates the long-run dependencies across the two dimensions of the matrix, similar to the framework recently proposed by \citet{li2024coint}.
More precisely, we impose a Kronecker structure on the $\balpha$, $\bbeta$, and $\bPhi_j$ terms of equation \eqref{eq:vecm} (see Section \ref{sec:MECM}).
This structure has three benefits over the traditional VECM.
First, it enables separate analysis of the cointegration dynamics across the rows and columns of the matrix, unlike the vectorized approach in equation \eqref{eq:vecm}.
We thus have a rank associated with the economic indicators ($r_1$, row rank) and a potentially different rank for the countries ($r_2$, column rank).
Second, we allow for a partial full-rank cointegrated system where only one of the two dimensions of the matrix-valued time series is rank-restricted.
Third, the resulting model is typically far more parsimonious, allowing for larger data sets than a traditional VECM.

While \citet{li2024coint} consider the MECM with fixed cointegration ranks mainly from a theoretical point of view, we complement their work by providing practitioners with practical tools, in the form of information criteria, to select the cointegration ranks $r_1$ and $r_2$ (see Section \ref{sec:estimationselection}).
Information criteria have been successfully applied in the cointegration literature \citep{Aznar2002selecting, cheng2009semiparametric} and can flexibly accommodate matrix-valued time series.
We demonstrate the good performance of the information criteria through a Monte Carlo simulation study in Section \ref{sec:simulation}.
Finally, our empirical results in Section \ref{sec:Application} reveal a single (restricted) long-run relationship between three economic indicators for the US, Germany, France, and Great Britain.
Replication material for the simulations and empirical analysis are available at \url{https://github.com/ivanuricardo/MECMrankdetermination}.

\section{The Matrix Error Correction Model}
\label{sec:MECM}

Let $\bY_t$ be an $N_1 \times N_2$ matrix-valued time series with I($1$) component series that follows the MECM($p$) model given by
\begin{align}
\label{eq:mecm}
    \bDelta \bY_t = \bD + \bU_1 \bU_3' \bY_{t-1} \bU_4 \bU_2' + \sum_{j=1}^{p} \bPhi_{1,j} \bDelta \bY_{t-j} \bPhi_{2,j}' + \bE_t,
\end{align}
where $\bD$ is the matrix of deterministic terms, $\bU_3 \in \mathbb{R}^{N_1 \times r_1}$ and $\bU_4 \in \mathbb{R}^{N_2 \times r_2}$ are the cointegrating matrices for the rows and columns of the matrix-valued time series, $\bU_1 \in \mathbb{R}^{N_1 \times r_1}$ and $\bU_2 \in \mathbb{R}^{N_2 \times r_2}$ are the corresponding adjustment coefficients, and $\bPhi_{1,j} \in \mathbb{R}^{N_1 \times N_1}$ and $\bPhi_{2,j} \in \mathbb{R}^{N_2 \times N_2}$ are the matrix autoregressive coefficients \citep{li2024coint}.\footnote{In the case of a stationary matrix AR($p$), different ranks are used for each matrix to distinguish between different right and left null space commonalities in the $\bU_i$, see \citet{hecq2024reduced}.}
We assume the errors follow a matrix-valued normal distribution \citep{dawid1981matrix}, namely
\begin{align*}
     \bE_t \sim MVN(\mathbf{0}, \bSigma_1, \bSigma_2) \Leftrightarrow \vecop{\bE_t} \sim N (\vecop{\mathbf{0}}, \bSigma_2 \otimes \bSigma_1),
\end{align*}
where $\bSigma_1 \in \mathbb{R}^{N_1 \times N_1}$ and $\bSigma_2 \in \mathbb{R}^{N_2 \times N_2}$ are positive definite matrices capturing the relations between the rows and columns of the matrix-valued errors, $MVN(\cdot,\cdot,\cdot)$ denotes the matrix-valued normal distribution, and $N(\cdot,\cdot)$ denotes the multivariate normal distribution.
Reorganizing equation \eqref{eq:mecm} to the traditional vector-valued set-up, where $\vecop{\bY_t} = \by_t$, gives the restricted VECM
\begin{align*}
    \bDelta \by_t = \bd + \underbrace{(\bU_2 \otimes \bU_1)}_{\balpha} \underbrace{(\bU_4 \otimes \bU_3)'}_{\bbeta'} \by_{t-1} + \sum_{j=1}^{p} \underbrace{(\bPhi_{2,j} \otimes \bPhi_{1,j})}_{\bPhi_j} \bDelta \by_{t-j} + \be_t.
\end{align*}
The MECM in equation \eqref{eq:mecm} thus implies a Kronecker structure on the adjustment coefficients $\balpha$, the cointegrating matrix $\bbeta$, and short-run coefficients $\bPhi_j$ of equation \eqref{eq:vecm}.

The imposed Kronecker structure puts restrictions on the coefficients by separating the cointegrating relations and adjustment coefficients across the two dimensions of the matrix-valued time series.
This separation not only enhances interpretability but also yields a substantial reduction in the number of parameters to be estimated.
The total number of effective parameters, excluding the constant term, is given by
\begin{equation}
\label{eq:numpars}
    \psi(r_1, r_2, p) = r_1 (2 N_1 - r_1) + r_2 (2 N_2 - r_2) + p (N_1^2 + N_2^2).
\end{equation}
As an example, consider a scenario where $(N_1, N_2) = (3, 4)$, $(r_1, r_2) = (1,1)$ and $p = 2$, then the MECM requires estimating $62$ parameters, whereas a comparable VECM with $N = 12$, $r = 1$, and $p = 2$ would require estimating $311$ parameters.

\begin{remark}
Model \eqref{eq:mecm} is not uniquely identifiable without additional restrictions on the parameters.
To resolve this, we impose that the top $r_1 \times r_1$ block of $\bU_3$ and the top $r_2 \times r_2$ block of $\bU_4$ are the identity matrix and set $\|\bSigma_1\|_F = 1$ such that $\bSigma_1$ is identified up to a sign change.
Identification restrictions on the short-run coefficient matrices $\bPhi_j$ are not required, but in case the dimension-specific matrices $\bPhi_{1,j}$ and $\bPhi_{2,j}$ are of interest, the same restriction may be used, namely, $\| \bPhi_{1,j}\|_F = 1$ for $j = 1, \dots, p$.
\end{remark}

\begin{remark}
We specify the short-run dynamics in equation \eqref{eq:mecm} as a matrix autoregression, following \citet{chen2021mar}.
It is, however, possible to replace this with unrestricted autoregressive components.
This substitution would increase the number of lagged autoregressive parameters from $ N_1^2 + N_2^2$ to $ N_1^2 N_2^2$.
\end{remark}

\begin{remark}
While the Kronecker product structure forms a natural way to reduce the dimensionality in MECMs for $N_1\times N_2$  matrix-valued data, note that it would be interesting to introduce a specification test to investigate whether the data support the presence of such a Kronecker product structure. To this end, an interesting avenue for future research would be to extend the specification test of  \cite{chen2021mar} for autoregressive models with matrix-valued time series to the MECM model set-up.
\end{remark}

\begin{remark}
We deliberately focus on MECMs of moderate dimension in this paper. 
The current MECM can, however, be extended to high-dimensional settings.
    Examples of possible extensions include applying techniques for stationary high-dimensional matrix-valued time series \citep{wang2019factor} to the cointegration case (e.g., nonstationary factor models as in \citealp{trapani2025inference}), as well as adapting methods for vector error correction models (e.g., sparse methods in \citealp{liao2015automated, wilms2016forecasting} or low-rank procedures in \citealp{cubadda2023vecim}) to the high-dimensional MECM set-up.
\end{remark}

\section{Estimation and Selection of Cointegration Ranks}
\label{sec:estimationselection}

The log-likelihood (up to a constant) of the MECM($p$) model in equation \eqref{eq:mecm} for fixed rank $r_1$ and $r_2$ is given by
\begin{align}
    \label{eq:loglike}
    \calL(\bTheta) = -\frac{T N_1}{2} \log |\bSigma_1| - \frac{T N_2}{2} \log |\bSigma_2| - \frac{1}{2}\sum_{t=1}^T \text{tr} ( \bSigma_1^{-1} (\bDelta \bY_t - \bU_1 \bU_3' \bY_{t-1} \bU_4 \bU_2' - \sum_{j=1}^p \bPhi_{1,j} \bDelta \bY_{t-j} \bPhi_{2,j}' - \bD) \nonumber \\
    \times \bSigma_2^{-1} (\bDelta \bY_t - \bU_1 \bU_3' \bY_{t-1} \bU_4 \bU_2' - \sum_{j=1}^p \bPhi_{1,j} \bDelta \bY_{t-j} \bPhi_{2,j}' - \bD)')
\end{align}
where $\bTheta$ collects all parameters.
The objective function is non-convex, but gradient descent can be used to solve problem \eqref{eq:loglike} in a computationally efficient way (see Appendix \ref{sec:gradientdescent}).

In practice, however, the ranks $r_1$ and $r_2$ are unknown and must be selected.
To this end, we use standard information criteria, namely, the Akaike Information Criterion (AIC, \citealp{akaike1974new}) and Bayesian Information Criterion (BIC, \citealp{schwarz1978estimating})
\begin{align*}
    \text{AIC}(r_1, r_2, p) &= -2 \calL(\widehat{\bTheta}) + 2 \psi(r_1, r_2, p), \\
    \text{BIC}(r_1, r_2, p) &= -2 \calL(\widehat{\bTheta}) + \ln(T) \psi(r_1, r_2, p),
\end{align*}
where $\calL(\widehat{\bTheta})$ is the value of the log-likelihood at the estimated parameters and $\psi(r_1, r_2, p)$ denotes the effective number of parameters (see eq. \ref{eq:numpars}).

\setlength{\tabcolsep}{1em}
\renewcommand{\arraystretch}{1.5}
\begin{table}[t]
  \centering
  \caption{MECM(0): Rank selection with AIC or BIC for $T = 100$ and $T = 250$ observations.}
  \label{tab:MECM0mat}
  \begin{tabular}{
    l
    >{\centering\arraybackslash}p{2.4cm}
    >{\centering\arraybackslash}p{2.4cm}
    >{\centering\arraybackslash}p{2.4cm}
    >{\centering\arraybackslash}p{2.4cm}
    >{\centering\arraybackslash}p{2.4cm}
  }
    \toprule
    \textbf{True Rank} & \textbf{Method} &\textbf{Average Ranks} & \textbf{Standard Deviation} & \textbf{Frequency Correct} \\
    \midrule
    (1,1) & AIC (100) & (1.02, 1.00) & (0.16, 0.00) & (0.98, 1.00)  \\
    & BIC (100) & (1.02, 1.00) & (0.13, 0.00) & (0.98, 1.00)  \\
    & AIC (250) & (1.01, 1.00) & (0.08, 0.00) & (0.99, 1.00)  \\
    & BIC (250) & (1.01, 1.00) & (0.04, 0.00) & (0.99, 1.00)  \\
    \hline
    (3,1) & AIC (100) & (3.00, 1.07) & (0.00, 0.26) & (1.00, 0.93)  \\
    & BIC (100) & (3.00, 1.01) & (0.00, 0.04) & (1.00, 0.99)  \\
    & AIC (250) & (3.00, 1.03) & (0.00, 0.17) & (1.00, 0.97)  \\
    & BIC (250) & (3.00, 1.00) & (0.00, 0.00) & (1.00, 1.00)  \\
    \hline
    (1,4) & AIC (100) & (1.13, 4.00) & (0.34, 0.03) & (0.87, 1.00) \\
    & BIC (100) & (1.06, 4.00) & (0.24, 0.03) & (0.94, 1.00) \\
    & AIC (250) & (1.03, 4.00) & (0.18, 0.00) & (0.97, 1.00) \\
    & BIC (250) & (1.02, 4.00) & (0.14, 0.00) & (0.98, 1.00) \\
    \hline
    (3,4) & AIC (100) & (3.00, 4.00) & (0.00, 0.00) & (1.00, 1.00) \\
    & BIC (100) & (3.00, 4.00) & (0.00, 0.00) & (1.00, 1.00) \\
    & AIC (250) & (3.00, 4.00) & (0.00, 0.00) & (1.00, 1.00) \\
    & BIC (250) & (3.00, 4.00) & (0.00, 0.00) & (1.00, 1.00) \\
    \bottomrule
  \end{tabular}
\end{table}

\section{Simulation Study}
\label{sec:simulation}

We conduct a simulation study to investigate the performance of our rank selection criteria.
The data-generating process (DGP) is examined under two scenarios.
The first is a MECM($0$), which omits the short-run dynamics from the model.
The second is a MECM($1$) with short-run dynamics.
Across all settings, we take $N_1 = 3$ and $N_2 = 4$ in line with our empirical application and generate $T+100$ observations from the MECM detailed in Appendix \ref{sec:signoise}, using the first $100$ as burn-in and taking $T=100$ and $T=250$.
The rank selection criteria then estimates MECMs across all possible combinations of the ranks ($r_1, r_2$), and selects the model with the lowest information criterion value.
We explore four settings of a reduced rank in the MECM: (i) fully reduced ($r_1 = r_2 = 1$), (ii) partially reduced first dimension ($r_1 = 1, r_2 = 4$), (iii) partially reduced second dimension ($r_1 = 3, r_2 = 1$), and (iv) no rank reduction ($r_1 = 3, r_2 = 4$).

To better understand the implications of these ranks in the context of our empirical application with $N_1=3$ economic indicators and $N_2=4$ countries, note that a rank of ($1,1$) indicates one cointegrating relation among the $N=12$ variables, suggesting that all indicators move together across different countries.
A rank of ($1,4$) reflects four distinct cointegrating relations, where all indicators co-move for each country separately.
A rank of ($3,1$) yields three cointegrating relations, where all countries co-move for each indicator separately.
Finally, with no rank reduction, the model captures a stationary process with no co-movements.

Table \ref{tab:MECM0mat} presents the results of the simulation study for the MECM($0$) DGP.
The results for the MECM($1$) DGP are similar and given in Table \ref{tab:MECM1mat} of Appendix \ref{sec:addsim}.
With a true rank of ($1,1$), all information criteria select the correct rank at a rate above $95\%$, with BIC performing best at a rate of over $95\%$.
As the number of observations grows, both AIC and BIC select the correct ranks more often.
Similar conclusions hold for cases with ranks ($3,1$) and ($1,4$).
AIC and BIC select the correct rank at least 85\% of the time with $T=100$, and it goes up to 100\% with $T=250$.
Under the full rank case, we always correctly select the ranks.

\section{Application}
\label{sec:Application}

\begin{figure}
    \centering
    \includegraphics[width=0.8\linewidth]{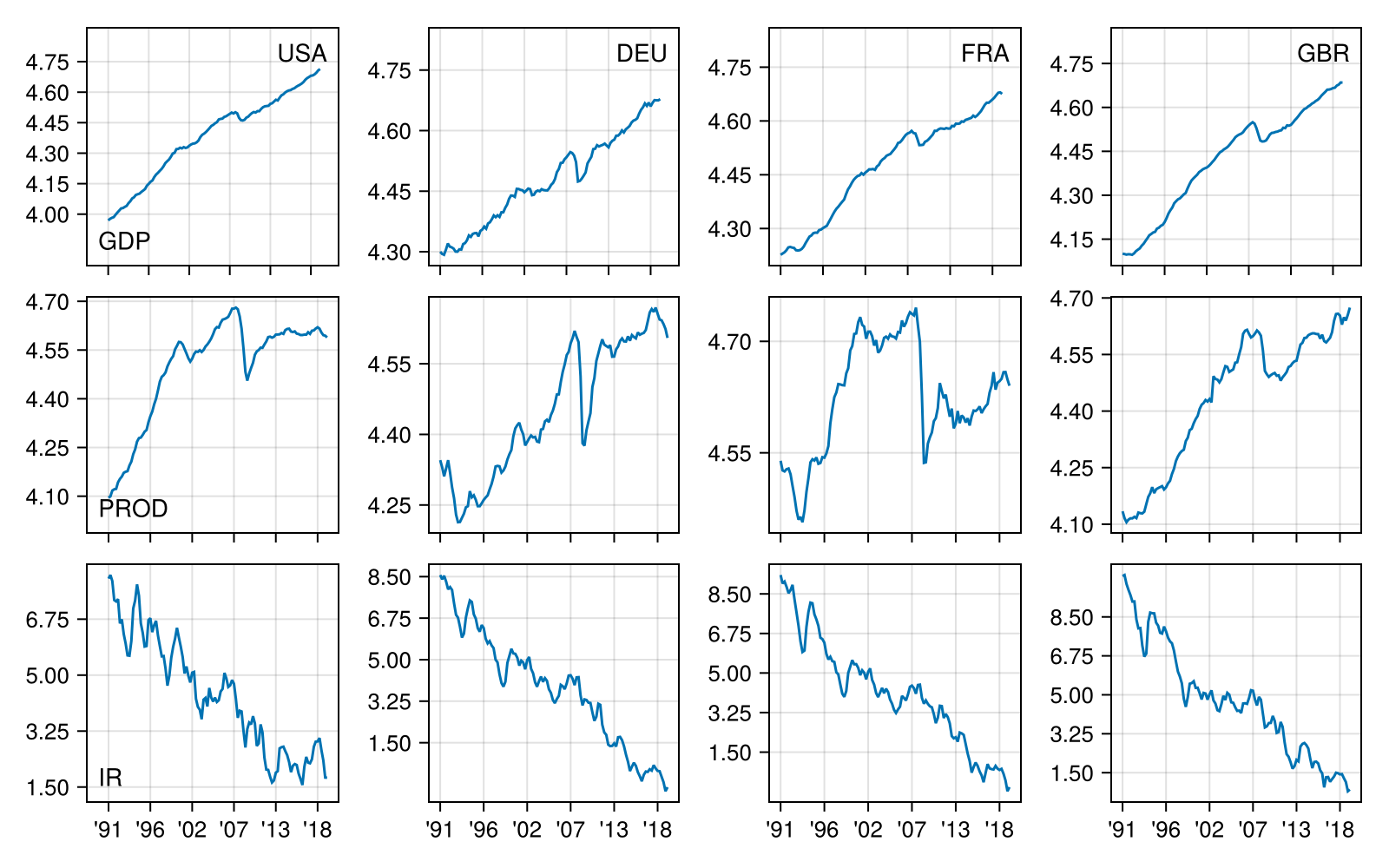}
    \caption{Time series plots for macroeconomic economic indicators (rows) and countries (columns).}
    \label{fig:globalplot}
\end{figure}

We consider quarterly macroeconomic data from 1991Q1 to 2019Q4 ($T=116$) on $N_1 = 3$ economic indicators across $N_2 = 4$ countries. 
This includes the log levels of real gross domestic product (GDP), the log levels of industrial production (PROD), and the levels of long-term interest rates for the United States (USA), Germany (DEU), France (FRA), and Great Britain (GBR).
The twelve time series are shown in Figure \ref{fig:globalplot}.

\begin{figure}
    \centering
    \includegraphics[width=0.65\linewidth]{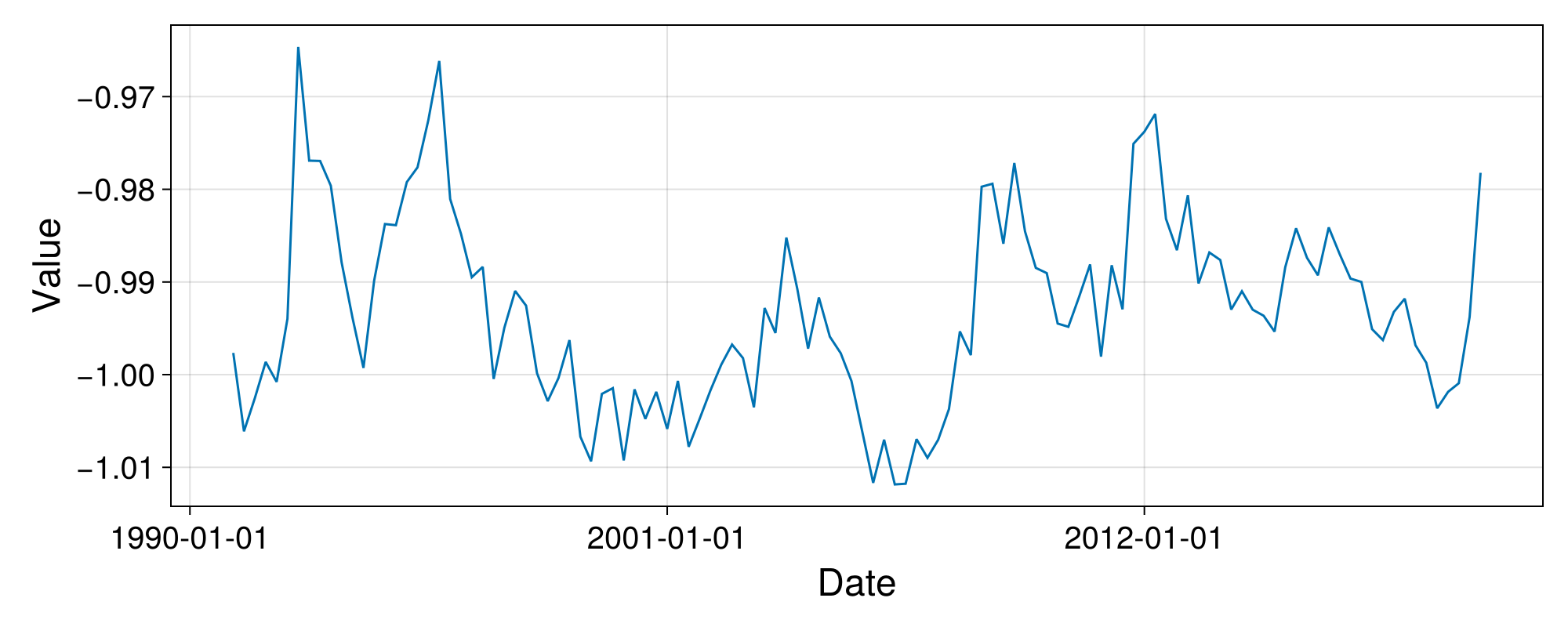}
    \caption{The cointegrated series $(\bU_4 \otimes \bU_3)' \vecop{\bY_{t-1}}$ capturing the long-run dynamics among the 12 variables.}
    \label{fig:cointvecs}
\end{figure}

All variables are found to be integrated of order I($1$) based on Augmented Dickey-Fuller tests.
We estimate a MECM($1$) for the rank selection criteria and both AIC and BIC select ($1,1$) as cointegrating ranks.
This implies one restricted cointegrating relation across all 12 variables, which is visualized in Figure \ref{fig:cointvecs} without adjustment for the short-run dynamics.
The corresponding estimated cointegrating matrices and adjustment coefficients are given in Table \ref{tab:estcoefficients}.

From the estimated indicator-specific cointegrating vector $\widehat{\bU}_3$, we observe that GDP positively co-moves with industrial production and long-term interest rates.
For the country-specific cointegrating vector $\widehat{\bU}_4$, the cointegrating relationship between the USA and France is stronger than that between the USA and either Germany or Great Britain.

\begin{table}[t]
    \centering
    \renewcommand{\arraystretch}{1.2}
    \setlength{\tabcolsep}{12pt}
    \begin{tabular}{lcc|lcc}
        \hline
        \multicolumn{3}{c|}{\textbf{Indicators}} & \multicolumn{3}{c}{\textbf{Countries}} \\
        \hline
        \rule{0pt}{3ex} 
        & $\widehat{\bU}_1$ & $\widehat{\bU}_3$ & & $\widehat{\bU}_2$ & $\widehat{\bU}_4$ \\
        \hline
        \textbf{GDP}  & -0.084 & 1.000 & \textbf{USA} & 0.088 & 1.000 \\
        \textbf{PROD} & -0.201 & -0.099 & \textbf{DEU} & 0.774 & 0.055 \\
        \textbf{IR}   & -7.843 & -0.021 & \textbf{FRA} & 1.103 & -1.092 \\
        & & & \textbf{GBR} & 0.674 & -0.196 \\
        \hline
    \end{tabular}
    \caption{Estimated parameter values for indicators and countries.}
    \label{tab:estcoefficients}
\end{table}

\bigskip
\noindent
{\bf Acknowledgements.}
We thank the editor and referee for their constructive comments which substantially improved the quality of the manuscript.
The last author was financially supported by the Dutch Research Council (NWO) under grant number VI.Vidi.211.032.

\bibliographystyle{asa}
\bibliography{updatedreferences}
\newpage

\appendix

\section{Gradient Ascent Algorithm}
\label{sec:gradientdescent}

We detail the gradient ascent algorithm \citep{wang2024high} for maximizing the log-likelihood given in equation \eqref{eq:loglike}.
We begin by initializing the $\bU_i$ for $i = 1, \dots, 4$ and $\bPhi_{1,j}, \bPhi_{2,j}$ via the projection method given in \cite{chen2021mar}.
$\bSigma_1$ and $\bSigma_2$ are initialized to be the identity, while $\bD$ is initialized to the zero matrix.
We then iterate over each parameter individually, using the previous parameters as inputs for the next gradient.
We hereby denote $\nabla \calL_x$  
as the partial gradient 
of parameter $x$.
Algorithm \ref{alg:gd} gives the steps.

\begin{algorithm}
\caption{Gradient ascent algorithm for MECM}
\label{alg:gd}
\begin{algorithmic}[0]
\State \textbf{Input}: \\
Ranks $(r_1, r_2)$, lag order $p$, initialization $\bU_i^{(0)}$ for $i = 1, \dots, 4$ and $\bSigma_i^{(0)}$ for $i = 1, 2$, step size $\eta_{\cdot}$, convergence tolerance $\epsilon = 0.01$,  maxiter = $500$.
\Repeat \quad s = 1, 2, \dots
    \State $\bD^{(s)} \gets \bD^{(s-1)} + \eta_{\bD} \nabla \mathcal{L}_{\bD}(\bD^{(s-1)})$
    \State $\bU_1^{(s)} \gets \bU_1^{(s-1)} + \eta_{\bU_1} \nabla \mathcal{L}_{\bU_1}(\bU_1^{(s-1)})$
    \State $\bU_3^{(s)} \gets \bU_3^{(s-1)} + \eta_{\bU_3} \nabla \mathcal{L}_{\bU_3}(\bU_3^{(s-1)})$
    \State normalize $\bU_3^{(s)}$
    \State $\bSigma_1^{(s)} \gets \bSigma_1^{(s-1)} + \eta_{\bSigma_1} \nabla \mathcal{L}_{\bSigma_1}(\bSigma_1^{(s-1)})$
    \State normalize $\bSigma_1^{(s)}$
    \State $\bU_2^{(s)} \gets \bU_2^{(s-1)} + \eta_{\bU_2} \nabla \mathcal{L}_{\bU_2}(\bU_2^{(s-1)})$
    \State $\bU_4^{(s)} \gets \bU_4^{(s-1)} + \eta_{\bU_4} \nabla \mathcal{L}_{\bU_4}(\bU_4^{(s-1)})$
    \State normalize $\bU_4^{(s)}$
    \State $\bSigma_2^{(s)} \gets \bSigma_2^{(s-1)} + \eta_{\bSigma_2} \nabla \mathcal{L}_{\bSigma_2}(\bSigma_2^{(s-1)})$
    \For{$j = 1, \dots, p$}
        \State $\bPhi_{1,j}^{(s)} \gets \bPhi_{1,j}^{(s-1)} + \eta_{\bPhi_{1,j}} \nabla \mathcal{L}_{\bPhi_{1,j}}(\bPhi_{1,j}^{(s-1)})$
        \State normalize $\bPhi_{1,j}^{(s)}$
        \State $\bPhi_{2,j}^{(s)} \gets \bPhi_{2,j}^{(s-1)} + \eta_{\bPhi_{2,j}} \nabla \mathcal{L}_{\bPhi_{2,j}}(\bPhi_{2,j}^{(s-1)})$
    \EndFor
\Until{$\mathcal{L}(\bTheta)^{(s)} - \mathcal{L}(\bTheta)^{(s-1)} < \epsilon$ or $s=\text{maxiter}$} 
\State \textbf{Return}: \quad $\bSigma_1^{(s)}, \bSigma_2^{(s)}, \bU_1^{(s)}, \bU_2^{(s)}, \bU_3^{(s)}, \bU_4^{(s)}, \bPhi_1^{(s)}, \bPhi_2^{(s)}, \bD^{(s)}$
\end{algorithmic}
\end{algorithm}

\section{MECM Reparameterization}
\label{sec:signoise}
We reparameterize the MECM in equation \eqref{eq:mecm} to a stable MAR \citep{li2024coint}.
The stable MAR reparameterization excluding the constant term is
\begin{align*}
   \binom{\boldsymbol{\beta}^{\prime} \boldsymbol{Y}_t}{\Delta \boldsymbol{Y}_t}= \bA \binom{\boldsymbol{\beta}^{\prime} \boldsymbol{Y}_{t-1}}{\Delta \boldsymbol{Y}_{t-1}}+\binom{\boldsymbol{\beta}^{\prime} \boldsymbol{E}_t}{\boldsymbol{E}_t}, \text { where } \bA :=\left(\begin{array}{cc}
    \left(\boldsymbol{I}_{r_1 \gamma_2}+\boldsymbol{\beta}^{\prime} \boldsymbol{\alpha}\right) & \boldsymbol{\beta}^{\prime} \boldsymbol{B} \\
    \boldsymbol{\alpha} & \boldsymbol{B}
\end{array}\right).
\end{align*}
Here, $\balpha = (\bU_2 \otimes \bU_1)$, $\bbeta = (\bU_4 \otimes \bU_3)$, and $\bB = (\bPhi_2 \otimes \bPhi_1)$.
The entries of the matrices $\bU_1, \bU_2, \bU_3,$ and $\bU_4$ are generated by sampling from the standard normal distribution and performing a QR decomposition to orthogonalize.
If the DGP has short-run dynamics, the same procedure is done with the MAR parameters $\bPhi_1$ and $\bPhi_2$.
Additionally, the error distribution is $\bE_t \sim MVN(\mathbf{0}, \bSigma_1, \bSigma_2)$, where $\bSigma_1$ and $\bSigma_2$ are diagonal with elements generated by fixing a signal-to-noise ratio of $0.7$.
We use the maximum eigenvalues of $\bA$ to determine the signal-to-noise ratio and to ensure we generate an I($1$) process in our simulation study.

\section{Additional Simulation Results}
\label{sec:addsim}

The results for the MECM($1$) are given in Table \ref{tab:MECM1mat}.
In the case with ranks ($1,1$) and $T=100$, the correct rank is selected in approximately $90\%$ of cases with the AIC, while BIC selects correctly over $99\%$ of cases.
When the number of observations is increased, the frequency of correctly selected ranks goes up to $97\%$ for AIC and $98\%$ for BIC.
In the case of partially full ranks and full ranks, we see a similar performance in the information criteria.
Finally, we investigated the case with errors from a multivariate normal distribution instead of a matrix-valued distribution and obtained similar findings.
Results are available upon request.

\setlength{\tabcolsep}{1em}
\renewcommand{\arraystretch}{1.5}
\renewcommand{\thetable}{C.\arabic{table}}
\setcounter{table}{0}
\begin{table}[ht]
  \centering
  \caption{MECM(1): Rank selection with AIC or BIC for $T = 100$ and $T = 250$ observations.}
  \label{tab:MECM1mat}
  \begin{tabular}{
    l
    >{\centering\arraybackslash}p{2.4cm}
    >{\centering\arraybackslash}p{2.4cm}
    >{\centering\arraybackslash}p{2.4cm}
    >{\centering\arraybackslash}p{2.4cm}
    >{\centering\arraybackslash}p{2.4cm}
  }
    \toprule
    \textbf{True Rank} & \textbf{Method} &\textbf{Average Rank} & \textbf{Std. Rank} & \textbf{Freq. Correct} \\
    \midrule
    (1,1) & AIC (100) & (1.06, 1.03) & (0.23, 0.21) & (0.94, 0.97)  \\
    & BIC (100) & (1.01, 1.00) & (0.08, 0.00) & (0.99, 1.00)  \\
    & AIC (250) & (1.06, 1.02) & (0.23, 0.15) & (0.94, 0.98)  \\
    & BIC (250) & (1.01, 1.00) & (0.08, 0.00) & (0.99, 1.00)  \\
    \hline
    (3,1) & AIC (100) & (3.00, 1.23) & (0.00, 0.59) & (1.00, 0.86)  \\
    & BIC (100) & (3.00, 1.01) & (0.00, 0.08) & (1.00, 0.99)  \\
    & AIC (250) & (3.00, 1.10) & (0.00, 0.40) & (1.00, 0.93)  \\
    & BIC (250) & (3.00, 1.01) & (0.00, 0.03) & (1.00, 0.99)  \\
    \hline
    (1,4) & AIC (100) & (1.01, 4.00) & (0.05, 0.00) & (0.99, 1.00) \\
    & BIC (100) & (1.01, 4.00) & (0.03, 0.00) & (0.99, 1.00) \\
    & AIC (250) & (1.00, 4.00) & (0.00, 0.00) & (1.00, 1.00) \\
    & BIC (250) & (1.00, 4.00) & (0.00, 0.00) & (1.00, 1.00) \\
    \hline
    (3,4) & AIC (100) & (3.00, 4.00) & (0.00, 0.00) & (0.99, 1.00) \\
    & BIC (100) & (3.00, 4.00) & (0.00, 0.00) & (0.99, 1.00) \\
    & AIC (250) & (3.00, 4.00) & (0.00, 0.00) & (1.00, 1.00) \\
    & BIC (250) & (3.00, 4.00) & (0.00, 0.00) & (1.00, 1.00) \\
    \bottomrule
  \end{tabular}
\end{table}

\end{document}